\documentclass[12pt,twoside]{article}
\usepackage[T1]{fontenc}
\usepackage[backend=bibtex,style=numeric-comp,autocite=plain,sorting=none,%
giveninits=true]{biblatex}
\usepackage{amsmath,amsfonts,amssymb,amsthm,mathabx}
\usepackage{times}
\usepackage{url}
\usepackage{tensor}
\usepackage{color}
\usepackage{caption}
\usepackage{enumitem}
\usepackage{graphicx}
\usepackage{array}
\usepackage{amsbsy}
\usepackage{appendix}
\usepackage{physics}
\usepackage{cancel}
\usepackage{yfonts}
\usepackage{chngcntr}
\usepackage{xcolor}
\usepackage{soul}
\usepackage{cancel}
\usepackage{setspace} 

\usepackage{hyperref}
\advance\textheight\topskip%
\def\scri{{\mathcal{J}}}

\begin{document}

\title{Thermodynamics of boosted Schwarzschild black holes}

\author{Glenn Barnich}


\def\mytitle{Thermodynamics of boosted Schwarzschild black holes}

\pagestyle{myheadings} \markboth{\textsc{\small G.~Barnich}} {
  Boosted Schwarzschild black holes thermodynamics}

\addtolength{\headsep}{4pt}

\begin{centering}

  \vspace{1cm}

  \textbf{\Large{\mytitle}}

    \vspace{1.5cm}

    {\large Glenn Barnich}

\vspace{.5cm}

\begin{minipage}{.9\textwidth}\small \it \begin{center}
    Theoretical and Mathematical Physics \\
    Universit\'e libre de Bruxelles and International Solvay Institutes\\
    Campus Plaine C.P. 231,
    B-1050 Brussels, Belgium \\
    E-mail: \href{mailto:Glenn.Barnich@ulb.be}{Glenn.Barnich@ulb.be}
    \end{center}
  \end{minipage}

\end{centering}

\vspace{1cm}
  
\begin{center}
  \begin{minipage}{.9\textwidth} \textsc{Abstract}. The equilibrium thermodynamics of
    boosted Schwarzschild black holes is worked out in a concise way and shown
    to agree with expectations from relativistic thermodynamics. 
 \end{minipage}
\end{center}

\vfill
\thispagestyle{empty}

\newpage

Whereas boosted black holes feature prominently in numerical relativity~\cite{Matzner:1998pt,Cook:2000vr}
and in more recent considerations on memory effects~\cite{Madler:2017umy,Madler:2017qlu}, their
equilibrium thermodynamics~\cite{Bardeen:1973gs,Carter1973,Carter:2009nex} does not seem to have been
developed in detail (see however~\cite{Horowitz:1997fr} for related considerations in a
different context). In this work, we focus on the simplest case of
four-dimensional asymptotically flat Schwarzschild black holes in order to
highlight the underlying mechanism.

While boosted Schwarzschild black holes are not new solutions to the Einstein
equations since they are obtained from the Schwarzschild solution through a
coordinate transformation, this is a ``large coordinate transformation/improper
gauge transformation''~\cite{teitelboim76:_lectur_gener_hamil_dynam,Benguria:1976in} that gives rise to additional non-vanishing
surface charges, in this case linear momentum. We provide a compact derivation
of how this changes the thermodynamics. Properties directly related to the
horizon (area, surface gravity, Smarr formula through Komar integral) are
unaffected by the boost.

The derivation is presented in Bondi form because the action of boosts is
particularly transparent at $\scri^+$. A derivation based on the ADM form will
yield identical results since for boosted Schwarzschild black holes, Bondi and
ADM four-momenta are identical due to the absence of radiation. The final result
is arguably appealing but ultimately not really surprising and probably known to
experts in the field -- it could have been inferred from Lorentz covariance.

{\bf Boosted Schwarzschild black holes}

We use mostly minus signature and $G=1=c$. Schwarzschild
black holes in outgoing Eddington-Finkelstein coordinates
$x^\mu=(u,r,x^A), x^A=(\zeta,\widebar\zeta)$, or equivalently in Bondi form, are described by
\begin{equation}
  \label{eq:4}
  \begin{split}
    & ds^2_M=-2U_Mdu^2+2dudr -r^2d^2\Omega,\\
    & U_M=-\frac{1}{2}+r^{-1}M,\quad d^2\Omega=\frac{2d\zeta d\widebar\zeta}{P_\circ^2},
      \quad P_\circ= \frac{|\zeta|^2+1}{\sqrt 2}.
  \end{split}
\end{equation}
At $M=0$, the line-element $ds_0$ corresponds to that of flat Minkowski
spacetime written in spherical (stereographic) coordinates with retarded time
$u=t-r$. A Newman-Penrose~\cite{Newman:1961qr} (co) tetrad associated to $ds^2_M$ is
\begin{equation}
  \label{eq:223}
  \begin{split}
    & l=\partial_r=du,\ n_M=\partial_u+U_M \partial_r=-U_M du+dr,\ m_\circ=r^{-1}P_\circ \widebar\partial=-\frac{r}{P_\circ}d\zeta,\\
    & ds^2_M=2ln_M-2m_\circ\widebar m_\circ,
  \end{split}
\end{equation}
which allows one to write the line-element in Kerr-Schild~\cite{Kerr:1965vyg} form as 
\begin{equation}
  \label{eq:7}
  ds^2_M=ds^2_0-2Mr^{-1}l^2.
\end{equation}

In order to describe boosts, one may start by a fractional linear transformation 
\begin{equation}
  \label{eq:5}
  \zeta =\frac{a\zeta^\prime+b}{c\zeta^\prime+d},\quad a,b,c,d\in \mathbb C,\quad ad-bc =1, 
\end{equation}
whose effect on the line element is 
\begin{equation}
  \label{eq:9}
  r^2d^2\Omega= r^2\frac{2d\zeta^\prime d\widebar\zeta^\prime}{P_g^2(x^\prime)}
  =r^2w_g^{-2}(x^\prime)d^2\Omega(x^\prime)
\end{equation}
where
\begin{equation}
  \label{eq:14}
  P_g(x)=\frac{|a\zeta+b|^{2}+|c\zeta +d|^{2}}{\sqrt 2},
                               \quad  w_g(x)=\frac{P_g(x)}{P_\circ(x)}= \frac{|a\zeta+b|^{2}+|c\zeta +d|^{2}}{|\zeta|^2+1}.
\end{equation}
In addition, one may use a class $\mathrm{III}$ Lorentz rotation to simplify the
complex dyad,
\begin{equation}
  \label{eq:12}
  \begin{split}
  & m_\circ=e^{\imath E_I}(x^\prime)m_g(x^\prime),\quad e^{\imath E_I}(x)=\frac{\widebar c\widebar\zeta+\widebar d}{c\zeta+d},\\
  & m_g(x^\prime)=r^{-1}P_g(x^\prime)\widebar\partial^\prime
    =-\frac{r}{P_g(x^\prime)}d\zeta^\prime.
  \end{split}
\end{equation}
After dropping the primes, the line element may thus be written as 
\begin{equation}
  \label{eq:13}
  ds^2_{M,g}=2ln_M-2m_g\widebar m_g.
\end{equation}
The rescaling of the radial coordinate accompanied by the opposite rescaling of
the retarded time coordinate,
\begin{equation}
  \label{eq:11}
  r =w_g(x^\prime)r^\prime,\quad u=w^{-1}_g(x^\prime)u^\prime, 
\end{equation}
brings the part of the line element involving the complex dyad back to the
standard spherical form.

The transformations~\eqref{eq:5} and~\eqref{eq:11} are the leading asymptotic part of a Lorentz
transformation. They can be completed by sub-leading corrections, ``pure gauge
transformations'', in at least two ways. The first is to express a Lorentz
transformation of Minkowski spacetime in curvilinear Bondi coordinates. The
Kerr-Schild form of a boosted Schwarzschild black hole is then obtained by
applying this coordinate transformation to~\eqref{eq:7} and using that, by construction,
this leaves $ds_0$ invariant. The second, which we will follow here, is to
complete this leading part into a Lorentz transformation inside the
$\mathrm{BMS}_4$ group at $\scri^+$. This will allow us to use earlier work on
how these transformations act on the data determining asymptotically flat
spacetimes in general, and on Schwarzschild black holes in particular, together
with suitable expressions for the associated surface charges.

For later use, note that (i) $w_g$ is the inverse of the boost factor used in
the context of spin and boost weighted fields~\cite{Held1970}
(see~\cite{Penrose:1984} for a review),
\begin{equation}
  \label{eq:15}
  K=e^{E_R}=w_g^{-1},
\end{equation}
(ii) it is of spin and boost weights $(0,1)$ and its expansion only involves the four
lowest harmonics,
  \begin{equation}
    \label{eq:6}
   w_g=c^{jm} {}_{0,1} Y_{jm}|_{{j\leq 1}},
 \end{equation}
 (iii) it is trivial in the case of a rotation, $w_g=1$ if
 $g\in \mathrm{SU}(2)\subset \mathrm{SL}(2,\mathbb C)$, and may be written as
 \begin{equation}
   \label{eq:16}
   w_g=u_0-u_i n^i,
 \end{equation}
 where $n^i$ is the unit radial vector while the future directed unit vector
 $u^a u_a=1$, $u_0(u_i)=\sqrt{1+|\vec u|^2}$, is parametrized in terms of boost
 or rapidity vectors $\vec \beta,\vec \omega$ as
  \begin{equation}
    \label{eq:243}
    \begin{split}
      u^0&=\frac{|a|^{2} + |c|^2+ |b|^{2}+|d|^{2}}{2}=\gamma=\cosh\omega,\\
      u^2&=\frac{\imath(a\widebar b+c\widebar d-b\widebar a-d\widebar c)}{2}=\gamma\beta^2=\sinh\omega\hat\omega^2,\\
      u^3&=\frac{|a|^{2}+|c|^{2}-|b|^{2}-|d|^{2}}{2}=\gamma\beta^3=
           \sinh\omega\hat\omega^3,\\
      u^1&=\frac{a\widebar b+c\widebar d
           +b\widebar a+d\widebar c}{2}=\gamma\beta^1=\sinh\omega\hat\omega^1.
    \end{split}
  \end{equation}

  {\bf Asymptotically flat data for boosted Schwarzschild black holes}

  In the context of locally asymptotically flat spacetimes at $\scri^+$ in the
  sense of Newman and Penrose~\cite{Newman:1961qr,newman:1980xx,Penrose:1986}, the free data that
  characterizes these solutions is given by the asymptotic part of the shear
  $\sigma^0$, the asymptotic parts of the Weyl components $\Psi^0_1,\Psi^0_2$ at fixed $u$,
  as well as $\Psi_0=O(r^{-5})$. A pedestrian approach to the $\mathrm{BMS}_4$
  group in this set-up consists in working out the combined coordinate
  transformations and frame rotations that preserve these solutions, together
  with its action on the free data that determines them~\cite{Barnich:2016lyg}.

  In particular, under a Lorentz transformation belonging to $\mathrm{BMS}_4$
  group at $\scri^+$, $\sigma^0$ transforms as a field of spin and boosts weights
  $(2,-1)$, while $\Psi^0_n$, $n=0,\dots 4$ transform as fields of spin and boost
  weights $(2-n,-3)$ corrected by terms involving $\Psi^0_m$ with $m> n$.

  For Schwarzschild black holes, the free data is
  \begin{equation}
    \label{eq:10}
    \sigma^0=0,\quad \Psi^0_2=-M,\quad \Psi^0_1=0,\quad \Psi_0=0.
  \end{equation}
  Under a Lorentz transformation corresponding to~\eqref{eq:5}, this data
  becomes
  \begin{equation}
    \label{eq:18}
    \sigma^{\prime 0}=0,\quad \Psi^{\prime 0}_2=-Mw_g^{-3},\quad \Psi^{\prime 0}_1=3M w_g^{-4} u^\prime \eth^\prime_\circ w_g,\quad \Psi^{\prime 0}_0=
    -6Mw_g^{-5}{(u^\prime \eth^\prime_\circ w_g )}^2,
  \end{equation}
  see~\cite{Barnich:2016lyg} and also~\cite{Barnich:2026wpw} for details.

  {\bf Bondi mass and angular momentum aspects, $\mathrm{BMS}_4$ charges}

  As a consequence of~\eqref{eq:18}, the Bondi mass aspect is
  \begin{equation}
    \label{eq:3}
    \Psi^\prime=Mw_g^{-3},
  \end{equation}
  while the angular momentum aspect vanishes,
  \begin{multline}
    \label{eq:17}
    \widebar\Psi^\prime_J(u^\prime)=-[\Psi^{\prime 0}_1
    +\frac{3}{2}\sigma^{\prime 0}\eth^\prime_\circ\widebar\sigma^{\prime 0}+\frac 12 \eth^\prime_\circ\widebar\sigma^{\prime 0}]-u^\prime\eth_\circ^\prime \Psi^\prime\\=-3M w_g^{-4} u^\prime \eth^\prime_\circ w_g-u^\prime\eth_\circ^\prime (Mw_g^{-3})=0.
  \end{multline}
  As a consequence, boosted Schwarzschild black holes have vanishing Lorentz
  charges, in line with the discussion in Appendix C of~\cite{Bonga:2018gzr}.

  Borrowing techniques from gravitational wave physics~\cite{Blanchet:1985sp}, a general
  expression for Bondi supermomentum has also been worked out
  in~\cite{Barnich:2026wpw}. Relevant for us here is only Bondi four-momentum
  which has been obtained by elementary means: for $k=2,3$, the integrals
        \begin{equation}
          \label{eq:58}
            \mathcal I_{k0}(u_0,u_i)=\int \frac{d^2\Omega}{4\pi}w_g^{-k}={(u_0)}^{-k}\int \frac{d^2\Omega}{4\pi}\frac{1}{{(1-\frac{u_i}{u_0}n^i)}^k},
          \end{equation}
         turn out to be,
        \begin{equation}
          \label{eq:325}
          \begin{split}
            \mathcal I_{20}(u_0,u_i)&=u_0^{-2}x^{-1}\frac{1}{2}\big[{(1-x)}^{-1}-{(1+x)}^{-1}]=u_0^{-2}\frac{1}{1-x^2},\quad x=\frac{|\vec u|}{u_0},\\
                            \mathcal I_{30}(u_0,u_i)&=u_0^{-3}x^{-1}\frac{1}{4}\big[{(1-x)}^{-2}-{(1+x)}^{-2}]=u_0^{-3}\frac{1}{{(1-x^2)}^2},
          \end{split}                                                      
        \end{equation}       
        so that
        \begin{equation}
          \label{eq:59}
          \mathcal I_{20}(u_0(u_i),u_i)=1.
        \end{equation}
It follows that Bondi mass, the monopole moment of the mass aspect, is 
\begin{equation}
    \label{eq:8}
      P_0=\int_{\mathbb S^2} \frac{d^2\Omega}{4\pi}M w_g^{-3}=M
      \mathcal I_{30}(u_0(u_i),u_i)=M u_0,
    \end{equation}
    while linear momentum, the dipole moment, is
    \begin{equation}
      \label{eq:1}
      P_i=\int_{\mathbb S^2} \frac{d^2\Omega}{4\pi}\Psi(-n_i)=-\frac 12 M \big[\partial_{u_i}\, \mathcal I_{20}(u_0,u_i)\big]\big|_{u_0(u_i)}=Mu_i,
    \end{equation}
with
        \begin{equation}
          \label{eq:76}
          P^2=P^a P_a=M^2,\quad P_a u^a= M\int_{\mathbb S^2} \frac{d^2\Omega}{4\pi}w_g^{-2}=M, 
        \end{equation}
in-line with expressions for ADM mass and linear momentum~\cite{Bonning:2003im}.         

{\bf Horizon area}

 In $u,r,\zeta,\bar\zeta$ coordinates, the horizon $\mathcal N$ is the null
  hypersurface defined by
  \begin{equation}
 U_M|_{\mathcal N}=0 \iff  r=2M. 
  \end{equation}
The area of the horizon is obtained by putting $r=2M,u=\mathrm{cte}$ and
        integrating over the volume using the induced metric, which yields
  \begin{equation}
    \label{eq:61}
 \mathcal A=16\pi M^2,
\end{equation}
on account of~\eqref{eq:59}, in-line with the analysis in~\cite{Akcay:2009owa}.

{\bf Surface gravity}

The (co) normal to the horizon is $l_{\mathcal N}=\tilde fdr=\tilde f(n+U_M l)$.
The horizon is a Killing horizon of $\xi=\partial_u=n-U_M l$ and, by using the spin
coefficients for the boosted Schwarzschild black holes (cf.~\cite{Barnich:2026wpw}), one finds
the standard result for the past horizon,
        \begin{equation}
          \label{eq:33}
          \xi\cdot \nabla \xi=-r^{-2}M(\xi+2U_M l) \Longrightarrow \kappa_{\mathcal N}=-\frac{1}{4M}. 
        \end{equation}

        {\bf Chemical potentials, fundamental relation and Massieu function from
          first law}

        Since the energy $P_0$ in~\eqref{eq:8} involves a relativistic factor due to the
        boost while the horizon physics is the same than for unboosted
        Schwarzschild black holes, the first law must be modified. In analogy
        with the first law for Kerr-Newman black holes, one expects
        \begin{equation}
          \label{eq:72}
          \delta E=T_H\delta S+\mu^i\delta P_i,\quad S=\frac{\mathcal A}{4}, 
        \end{equation}
        where the Hawking temperature $T_H$ is proportional to minus the surface
        gravity $-\kappa_{\mathcal N}$ of the past horizon and $S$ is the entropy of
        the black hole solutions, while $\mu^i$ are suitable chemical potentials
        for linear momentum. Given the expressions~\eqref{eq:8},~\eqref{eq:1} for $E=P_0,P_i$ and
        the area~\eqref{eq:61}, this may be solved to find the temperature and chemical
        potentials,
        \begin{equation}
          \label{eq:73}
           T_H={(8\pi M u_0)}^{-1},\quad \mu^i=-\frac{u^i}{u_0}. 
         \end{equation}
         Accordingly, boosted black holes are colder than those at rest, in
         agreement with the Planck-Einstein transformation formula.
         
        In terms of $\beta^0=\frac{1}{T_H}$, $\beta^i=-\frac{\mu^i}{T_H}$, the
        first law may then be re-written in a special-relativistic form as
        \begin{equation}
          \label{eq:74}
      \boxed{  \delta S=\beta^a\delta P_a, \quad \beta^a=8\pi M u^a=8\pi P^a},
      \end{equation}
      with associated fundamental relation
      \begin{equation}
          \label{eq:22}
          \boxed{S(P^a)=4\pi P^a P_a}, 
        \end{equation}
        and Massieu function
        \begin{equation}
          \label{eq:30}
       \boxed{  \Phi(\beta^a)}=(S-\beta^a P_a)|_{P_a=P_a(\beta^b)}\boxed{=-\frac{1}{16\pi}\beta^a\beta_a}, \quad \frac{\partial \Phi}{\partial \beta^b}=-P_b.
       \end{equation}

       {\bf Geometrical derivation of first law}

        What remains to be done to fully validate the first law in the sense of
        black hole mechanics is to derive the surface gravity
        $\kappa_{\mathcal N}=-{(16\pi^2 M u_0)}^{-1}$ and the chemical potentials $\mu^i$ from
        the geometry of the horizon understood as a suitable Killing horizon.

        To do so, we use that the Bondi charges for boosted Schwarzschild black
        holes are ``integrable'' at $\scri^+$ (the surface integral term is a
        variation, cf.~\cite{Regge:1974zd}, no $\Theta$ term in the language of Wald and
        Zoupas~\cite{Wald:1999wa}). In this Newman-Penrose context, the expression for the
        variation of the BMS charges has been worked out for instance
        in~\cite{Barnich:2019vzx}. For boosted Schwarzschild black holes, it reduces to that of
        the supertranslation charges
        \begin{equation}
          \label{eq:19}
          \delta Q_T=-\delta \int \frac{d^2\Omega}{8\pi} T (\Psi^0_2+{\rm c.c.})=\delta
          \int \frac{d^2\Omega}{4\pi} T Mw_g^{-3}. 
        \end{equation}
        The variation of the Bondi mass is obtained when using $T=1$
        correponding to the vector $\partial_{u^\prime}$. Since to leading order, the
        relation to the Killing vector $\xi=\partial_u$ is determined according to~\eqref{eq:11} by
        $u^\prime=w_g u$ so that $\partial_u=w_g\partial_{u^\prime}=(u_0-u_i n^i)\partial_{u^\prime}$. Hence, in
        that case, $T=w_g$. In other words, when using~\eqref{eq:59}, $Q_{w_g}=M$ on the
        one hand, while~\eqref{eq:8} and~\eqref{eq:1} imply that
        $Q_{w_g}= u^aP_a$ on the other. For the variation, one has
        $\delta Q_{w_g}=u^a\delta P_a$ since $\delta u^a u_a=0$, and also,
        when using for instance the general arguments of~\cite{Iyer:1994ys} to evaluate the variation at the horizon, that
        \begin{equation}
          \label{eq:2}
          \delta Q_{w_g}=-\frac{\kappa_{\mathcal N}}{2\pi}\delta \frac{\mathcal A}{4}. 
        \end{equation}
        
        Accordingly, the first law (for past horizons taking orientation into
        account) is given by
        \begin{equation}
          \label{eq:34}
          -\frac{\kappa_{\mathcal N}}{2\pi}\delta \frac{\mathcal A}{4}=u^0\delta P_0+u^i\delta P_i \iff
          \delta S=\frac{2\pi u^0}{-\kappa_{\mathcal N}}\delta P_0+\frac{2\pi u^i}{-\kappa_{\mathcal N}}\delta P_i,
        \end{equation}
        in agreement with~\eqref{eq:74}.

        {\bf Outlook}

        From the viewpoint of the Poincar\'e charges, a Schwarzschild black hole
        at fixed mass corresponds to a massive spinless representation of the
        Poincar\'e group. Equivalently, its Pauli-Lubanski vector vanishes
        identically, $W^a=0$. The fundamental relation for the entropy $S(P_a)$
        in~\eqref{eq:22} shows that the entropy is a Poincar\'e invariant, depending only
        on the unique non-trivial Casimir $P^2$.

        More generally, in the absence of radiation, $\mathrm{BMS}_4$ surface
        charges transform in the coadjoint representation,
        \begin{equation}
          \label{eq:20}
          \begin{split}
            \mathcal J^\prime(x^\prime)&=e^{-3E_R+\imath E_I}(x^\prime)\Big(\mathcal J+\big(\frac 12 \mathcal T\widebar\eth_\circ\mathcal P+\frac{3}{2}\widebar\eth_\circ \mathcal T\mathcal P\big)\Big)(x),\\
            \widebar{\mathcal J}^\prime(x^\prime)&=e^{-3E_R-\imath E_I}(x^\prime)\Big(\widebar{\mathcal J}+\big(\frac 12 \mathcal T\eth_\circ\mathcal P+\frac{3}{2}\eth_\circ \mathcal T\mathcal P\big)\Big)(x),\\
            \mathcal P^\prime(x^\prime)&=e^{-3E_R}(x^\prime)\mathcal P(x),
          \end{split}
        \end{equation}
        see~\cite{Barnich:2021dta} for details. Accordingly, acting with (super)-translations on
        Schwarzschild black holes ($\mathcal J=0=\widebar{\mathcal J}$),
        generates a non-trivial orbital contribution to the Lorentz charges,
        while the intrinsic spin remains zero. Likewise, for Kerr black holes, which carry 
        non-vanishing intrinsic spin, boosts and rotations will affect the value of
        the Lorentz charges inside a given BMS coadjoint orbit. It should be instructive to generalize the
        considerations here by working out in detail the thermodynamics of
        $\mathrm{BMS}_4$ transformed Kerr black holes.

        Ultimately, it would of course be desirable to recover the Massieu
        function in~\eqref{eq:30} as the thermodynamic limit of the logarithm of a
        relativistic partition function
        \begin{equation}
          \label{eq:21}
          Z(\beta^a)={\rm Tr}\ e^{-\beta^a \hat P_a}
        \end{equation}
        defined on an appropriate Hilbert space.
        
        {\bf Acknowledgements}

        The author is grateful to R.~Emparan for drawing his attention to
        reference~\cite{Horowitz:1997fr}. This work is supported by the F.R.S.-FNRS Belgium
        through convention IISN 4.4516.26 and CDR grant J.0006.26.

\addcontentsline{toc}{section}{References}

\printbibliography%

\end{document}